\documentclass[twocolumn]{aastex631}
\usepackage[utf8]{inputenc}

\begin{document}
\maxdeadcycles=200

\newcommand{\LyA}{Lyman-$\alpha$}
\newcommand{\Mgii}{\ion{Mg}{2}}
\newcommand{\EWFL}{W$_0^{\lambda 2796}$}
\newcommand{\kms}{${\rm km\,s}^{-1}$}
\newcommand{\voff}{${v}_{\rm off}$}
\newcommand{\avgebv}{$\overline{\textrm{E(B-V)}}$}

\newcommand{\LGN}[1]{\textcolor{blue}{\bf #1}}
\newcommand{\citehere}{\textcolor{red}{\bf Citation Needed}} 

\title{DESI \Mgii\ Absorbers: Extinction Characteristics \& Quasar Redshift Accuracy}

\author[0000-0002-5166-8671]{Lucas Napolitano}
\author{Adam D.\ Myers}
\affiliation{Department of Physics \& Astronomy, University of Wyoming, 1000 E. University, Dept. 3905, Laramie, WY 82071, USA}
\author[0000-0003-1251-532X]{Victoria A. Fawcett}
\affiliation{School of Mathematics, Statistics and Physics, Newcastle University, Newcastle upon Tyne NE1 7RU, UK}

\author{Jessica Aguilar}
\affiliation{Lawrence Berkeley National Laboratory, 1 Cyclotron Road, Berkeley, CA 94720, USA}
\author[0000-0001-6098-7247]{Steven Ahlen}
\affiliation{Physics Dept., Boston University, 590 Commonwealth Avenue, Boston, MA 02215, USA}
\author[0000-0001-9712-0006]{Davide Bianchi}
\affiliation{Dipartimento di Fisica ``Aldo Pontremoli'', Universit\`a degli Studi di Milano, Via Celoria 16, I-20133 Milano, Italy}
\author{David Brooks}
\affiliation{Department of Physics \& Astronomy, University College London, Gower Street, London, WC1E 6BT, UK}
\author{Todd Claybaugh}
\affiliation{Lawrence Berkeley National Laboratory, 1 Cyclotron Road, Berkeley, CA 94720, USA}
\author[0000-0002-5954-7903]{Shaun Cole}
\affiliation{Institute for Computational Cosmology, Department of Physics, Durham University, South Road, Durham DH1 3LE, UK}
\author[0000-0002-1769-1640]{Axel de la Macorra}
\affiliation{Instituto de F\'{\i}sica, Universidad Nacional Aut\'{o}noma de M\'{e}xico,  Cd. de M\'{e}xico  C.P. 04510,  M\'{e}xico}
\author[0000-0002-5665-7912]{Biprateep Dey}
\affiliation{Department of Astronomy \& Astrophysics, University of Toronto, Toronto, ON M5S 3H4, Canada}
\affiliation{Department of Physics \& Astronomy and Pittsburgh Particle Physics, Astrophysics, and Cosmology Center (PITT PACC), University of Pittsburgh, 3941 O'Hara Street, Pittsburgh, PA 15260, USA}
\author[0000-0002-3033-7312]{Andreu Font-Ribera}
\affiliation{Department of Physics \& Astronomy, University College London, Gower Street, London, WC1E 6BT, UK}
\affiliation{Institut de F\'{i}sica d’Altes Energies (IFAE), The Barcelona Institute of Science and Technology, Campus UAB, 08193 Bellaterra Barcelona, Spain}
\author[0000-0002-2890-3725]{Jaime E. Forero-Romero}
\affiliation{Departamento de F\'isica, Universidad de los Andes, Cra. 1 No. 18A-10, Edificio Ip, CP 111711, Bogot\'a, Colombia}
\affiliation{Observatorio Astron\'omico, Universidad de los Andes, Cra. 1 No. 18A-10, Edificio H, CP 111711 Bogot\'a, Colombia}
\author{Enrique Gaztañaga}
\affiliation{Institut d'Estudis Espacials de Catalunya (IEEC), 08034 Barcelona, Spain}
\affiliation{Institute of Cosmology and Gravitation, University of Portsmouth, Dennis Sciama Building, Portsmouth, PO1 3FX, UK}
\affiliation{Institute of Space Sciences, ICE-CSIC, Campus UAB, Carrer de Can Magrans s/n, 08913 Bellaterra, Barcelona, Spain}
\author[0000-0003-3142-233X]{Satya Gontcho A Gontcho}
\affiliation{Lawrence Berkeley National Laboratory, 1 Cyclotron Road, Berkeley, CA 94720, USA}
\author{Gaston Gutierrez}
\affiliation{Fermi National Accelerator Laboratory, PO Box 500, Batavia, IL 60510, USA}
\author[0000-0002-6550-2023]{Klaus Honscheid}
\affiliation{Center for Cosmology and AstroParticle Physics, The Ohio State University, 191 West Woodruff Avenue, Columbus, OH 43210, USA}
\affiliation{Department of Physics, The Ohio State University, 191 West Woodruff Avenue, Columbus, OH 43210, USA}
\affiliation{The Ohio State University, Columbus, 43210 OH, USA}
\author{Stephanie Juneau}
\affiliation{NSF NOIRLab, 950 N. Cherry Ave., Tucson, AZ 85719, USA}
\author{Andrew Lambert}
\affiliation{Lawrence Berkeley National Laboratory, 1 Cyclotron Road, Berkeley, CA 94720, USA}
\author[0000-0003-1838-8528]{Martin Landriau}
\affiliation{Lawrence Berkeley National Laboratory, 1 Cyclotron Road, Berkeley, CA 94720, USA}
\author[0000-0001-7178-8868]{Laurent Le Guillou}
\affiliation{Sorbonne Universit\'{e}, CNRS/IN2P3, Laboratoire de Physique Nucl\'{e}aire et de Hautes Energies (LPNHE), FR-75005 Paris, France}
\author[0000-0002-1125-7384]{Aaron Meisner}
\affiliation{NSF NOIRLab, 950 N. Cherry Ave., Tucson, AZ 85719, USA}
\author{Ramon Miquel}
\affiliation{Instituci\'{o} Catalana de Recerca i Estudis Avan\c{c}ats, Passeig de Llu\'{\i}s Companys, 23, 08010 Barcelona, Spain}
\affiliation{Institut de F\'{i}sica d’Altes Energies (IFAE), The Barcelona Institute of Science and Technology, Campus UAB, 08193 Bellaterra Barcelona, Spain}
\author[0000-0002-2733-4559]{John Moustakas}
\affiliation{Department of Physics and Astronomy, Siena College, 515 Loudon Road, Loudonville, NY 12211, USA}
\author[0000-0001-8684-2222]{Jeffrey A. Newman}
\affiliation{Department of Physics \& Astronomy and Pittsburgh Particle Physics, Astrophysics, and Cosmology Center (PITT PACC), University of Pittsburgh, 3941 O'Hara Street, Pittsburgh, PA 15260, USA}
\author[0000-0001-7145-8674]{Francisco Prada}
\affiliation{Instituto de Astrof\'{i}sica de Andaluc\'{i}a (CSIC), Glorieta de la Astronom\'{i}a, s/n, E-18008 Granada, Spain}
\author[0000-0001-6979-0125]{Ignasi P\'erez-R\`afols}
\affiliation{Departament de F\'isica, EEBE, Universitat Polit\`ecnica de Catalunya, c/Eduard Maristany 10, 08930 Barcelona, Spain}
\author{Graziano Rossi}
\affiliation{Department of Physics and Astronomy, Sejong University, Seoul, 143-747, Korea}
\author[0000-0002-9646-8198]{Eusebio Sanchez}
\affiliation{CIEMAT, Avenida Complutense 40, E-28040 Madrid, Spain}
\author{David Schlegel}
\affiliation{Lawrence Berkeley National Laboratory, 1 Cyclotron Road, Berkeley, CA 94720, USA}
\author{Michael Schubnell}
\affiliation{Department of Physics, University of Michigan, Ann Arbor, MI 48109, USA}
\affiliation{University of Michigan, Ann Arbor, MI 48109, USA}
\author{David Sprayberry}
\affiliation{NSF NOIRLab, 950 N. Cherry Ave., Tucson, AZ 85719, USA}
\author[0000-0003-1704-0781]{Gregory Tarl\'{e}}
\affiliation{University of Michigan, Ann Arbor, MI 48109, USA}
\author{Benjamin Alan Weaver}
\affiliation{NSF NOIRLab, 950 N. Cherry Ave., Tucson, AZ 85719, USA}
\author[0000-0002-6684-3997]{Hu Zou}
\affiliation{National Astronomical Observatories, Chinese Academy of Sciences, A20 Datun Rd., Chaoyang District, Beijing, 100012, P.R. China}

\begin{abstract}
In this paper, we study how absorption-line systems affect the spectra and redshifts of quasars (QSOs), using catalogs of \Mgii\ absorbers from the early data release (EDR) and first data release (DR1) of the Dark Energy Spectroscopic Instrument (DESI). We determine the reddening effect of an absorption system by fitting an un-reddened template spectrum to a sample of 50{,}674 QSO spectra that contain \Mgii\ absorbers. We find that reddening caused by intervening absorbers (\voff\ $>3500$ \kms) has an average color excess of \avgebv\ = 0.04 magnitudes. We find that the E(B-V) tends to be greater for absorbers at low redshifts, or those having \Mgii\ absorption lines with higher equivalent widths, but shows no clear trend with \voff\ for intervening systems. However, the \avgebv\ of associated absorbers, those at  \voff\ $<3500$ \kms, shows a strong trend with \voff, increasing rapidly with decreasing \voff\ and peaking ($\sim$0.15 magnitudes) around \voff\ $=0$ \kms. We demonstrate that \Mgii\ absorbers impact redshift estimation for QSOs by investigating the distributions of \voff\ for associated absorbers. We find that at z $>1.5$, these distributions broaden and bifurcate in a nonphysical manner. In an effort to mitigate this effect, we mask pixels associated with the \Mgii\ absorption lines and recalculate the QSO redshifts. We find that we can recover \voff\ populations in better agreement with those for z $<1.5$ absorbers and in doing so typically shift background QSO redshifts by $\Delta \textrm{z} \approx \pm\ 0.005$.

\end{abstract}

\section{Introduction}
Since their discovery in the early 1960s, \citep[e.g.][]{QSODiscoveryI,QSODiscoveryII} quasars, also known as quasi-stellar objects (QSOs), have been a cornerstone of cosmological research, and their spectra have been collected in large number \citep[e.g.][]{SDSS7,SDSS12,SDSS16}. However, insight into the host galaxy environments of QSOs has been challenging to acquire. This challenge arises partly because QSOs have luminosities that can be vastly higher than those of their host galaxies, \citep[e.g.][]{QSOHost} and also because the distance to these systems makes the galaxy and quasar appear as a single point source. The use of absorption line systems (ALS) \citep[e.g.][]{Bahcall1965,Bahcall1966A,Bahcall1966B}, specifically associated absorption systems (AAS), circumvents this challenge by measuring an absorption signal, and is therefore able to provide information on the physical conditions of QSO host galaxies \citep[e.g][]{VandenBerk2001,QSOHostProfiles,COSHalos}.

Results from simulations \citep[e.g.][]{AASSim}, theory \citep[e.g.][]{AASTheory}, and observation \citep[e.g.][]{York2008,ShenMenard2013,AASObs}, are all roughly in agreement that AAS are found within a velocity offset of $\pm3500$ ${\rm km\,s}^{-1}$ relative to the QSO host galaxy, where velocity offset is parameterized in terms of the absorption line system and QSO redshifts as:
\begin{equation}
    \rm{v}_{\rm off} = c \frac{z_{\rm QSO}-z_{\rm ALS}}{1+z_{\rm QSO}}
\end{equation}
Multiple explanations have been proposed for the origin of AAS, including, as outlined in \citet{ShenMenard2013}: (1) absorption by material in the immediate ($\lesssim 200$ pc) vicinity of the QSO \citep[e.g][]{Hamann1995,BarlowSargent1997}, (2) absorption by material accelerated by starburst shocks \citep[e.g][]{Heckman1990,FuStockton2007}, (3) absorption by halo clouds in the QSO host galaxy \citep[e.g][]{Heckman1991,Chelouche2008} or (4) absorption by clouds in external galaxies clustered around the QSO  \citep[e.g][]{Weymann1979,Wild2008}. It is probable that the population of AAS is composed of systems created by a combination of these scenarios.

Absorbers with \voff\ $>3500$ ${\rm km\,s}^{-1}$ are considered to be intervening absorption systems (IAS). The absorption in these cases results from dust and gas in a galaxy that is not believed to be physically associated with the QSO, but instead one that lies along the same line of sight \citep[e.g.][]{York2008,ShenMenard2013,Khare2014,Chen2020,Abhijeet2021}. QSO redshifts are typically determined using broad emission lines, and as such, are subject to large inherent uncertainties \citep[e.g.][]{SDSS12}. These uncertainties impact the robustness of a velocity offset value, and as such, different values for the separation of associated and intervening absorption systems have been quoted throughout the literature \citep{ShenMenard2013}. By investigating the variance of an absorber's reddening effect with its velocity offset, we aim to support and/or refine the canonical delineating value of $3500$ ${\rm km\,s}^{-1}$.

Previous studies have shown, by fitting templates to composite spectra, that QSOs with AAS have significantly more dust extinction than IAS QSOs, and in both classes of absorbers, the extinction curve has been shown to be similar to that of the Small Magellanic Cloud (SMC) \citep[e.g.][]{VandenBerk2008,ShenMenard2013,Khare2014,Chen2020}. In this paper we will study the extinction of both IAS and AAS observed by the Dark Energy Spectroscopic Instrument, \citep[DESI; e.g.][]{Snowmass2013,DESI2016a.Science,DESI2022.KP1.Instr,DESIEDR,DESIFocalPlane,DESICorrector,DESI2023a.KP1.SV} which is currently releasing the results of its first year of analysis \citep[e.g][]{DESI2024.II.KP3,DESI2024.III.KP4,DESI2024.IV.KP6,DESI2024.V.KP5,DESI2024.VI.KP7A,DESI2024.VII.KP7B,DESI2024.I.DR1} Rather than using composite spectra of absorbed QSOs, we will instead fit an extinction curve to each individual absorbed spectrum, using a blue QSO composite derived in \citet{Fawcett2022} as a comparison.

This approach will allow us to use the large number of IAS and AAS available from the Survey Validation phase and Year 1 data release of DESI \citep[e.g.][]{DESI2022.KP1.Instr,DESI2023a.KP1.SV,DESIEDR} in order to investigate the extinction characteristics of these systems and their dependence on the \voff, redshift and equivalent width (\EWFL) of the absorbers. We will further use the extinction results to investigate the separation in \voff\ of IAS and AAS. Finally, we will consider the effect of AAS on the redshifts of DESI QSOs, as we anticipate that absorption systems may affect the ability of DESI to accurately determine background QSO redshifts.

This paper will be organized in the following way: In \S2 we will discuss our input absorber catalog, as well as the methods by which we fit extinction curves to our data. In \S3 we will present our results, focusing on the variation of fit color excess values with redshift and \voff. In \S4 we will discuss the physical implications of these results, as well as their possible effects on QSO redshifts. We present our conclusions in \S5.

\section{Data and Methods}
In this section we will discuss the characteristics of our input catalog of absorption systems. We will then detail the process by which we fit the extinction of QSO spectra with \Mgii\ absorbers.

\subsection{\Mgii\ Absorber Catalog}
The methods by which we have constructed the catalog of \Mgii\ absorbers we use in this study are detailed in \citet{MgIICat}. To summarize, absorbers are detected using a line-fitting process and then line properties are fit using a Markov Chain Monte Carlo (MCMC) sampler. The catalog presented in \citep{MgIICat} utilized only DESI EDR spectra. This work will additionally use the results of a newly released DR1 \Mgii\ absorber catalog created using the same methods as its EDR counterpart \citep[][]{DESI2024.I.DR1}.

Both the EDR and DR1 catalogs of absorbers are constructed from input QSO catalogs, whose specifications and methodologies are detailed in \citet{QSOTS}. In determining QSO redshifts, initial values are calculated as part of the main spectroscopic pipeline by {\tt Redrock} \citep[RR;][]{PipelinePaper,RR-Abhijeet,RedRockPaper}\footnote{\url{https://github.com/desihub/redrock}} which uses chi2 minimization of PCA templates to determine the redshift of a given spectrum. Two afterburner programs, {\tt QuasarNet} \citep{QN}\footnote{\url{https://github.com/ngbusca/QuasarNET}}, a neural network based QSO redshift fitter \citep[see also][]{Farr2020}, and a \Mgii-emission-based code designed to identify low-z QSOs are then run. The results of these afterburners can inform the need for a change in object type, or an additional {\tt Redrock} run with adjusted redshift priors, however the final redshift value is always determined by {\tt Redrock}. A description of the cuts used in determining DESI's QSO sample, as well as the methodology used in determining these cuts, can be found in \citet{QSOVI}.

Briefly detailing the statistics of the absorber catalogs, from the EDR sample we detect 23{,}921 absorbers from a parent sample of 83{,}207 QSOs. From the DR1 sample, we detect an additional 270{,}529 absorbers in 1.47 million QSO spectra. We note that the ratio of absorbers detected relative to the QSO sample has decreased, which we attribute to lower exposure times and fewer QSOs with multiple observations in DR1 compared to EDR \citep{SVOPS,DESI2023a.KP1.SV}. 

In particular, we find a dearth of QSOs with detected absorbers at redshifts $1.6<z<2.15$ when comparing the EDR and DR1 samples. We attribute this difference to a change in targeting philosophy: during EDR all QSOs at $z>1.6$ received 4 high-priority observations, whereas only \LyA\ QSOs at $z>2.15$ are intended to receive 4 observations at high priority during DESI's main survey \citet{DESIEDR}. With future DESI releases, we anticipate the ratio of absorbers detected relative to the QSO sample to increase as additional observations of QSOs are taken and their data coadded.

In order to reduce the sample of detected absorbers to one suitable for extinction fitting, we make some considerations. There exist a number of QSO spectra in our sample in which we detect multiple \Mgii\ absorbers; each of these absorbers will redden the background QSO spectra and do so with differing intensity across all wavelengths. The individual effect of these absorbers will be highly degenerate with each other. With this in mind, we choose to fit only those systems with a single detected absorber.

We further consider the possibility that for absorbers at large physical separations from their background QSO, the spectrum could be reddened by interactions with the intergalactic medium or absorption systems we are unable to detect, such as those at redshifts $z>2.5$ or those with weak \Mgii\ absorption lines \citep[see \S2.4.2 of][]{MgIICat}. With this in mind we consider only those absorbers with \voff\ $ < 20{,}000$ \kms. This will enable us to consider the separation between IAS and AAS in detail, as well as determine the average values of color excess for these classes of absorbers.

\begin{figure}[ht!]
\plotone{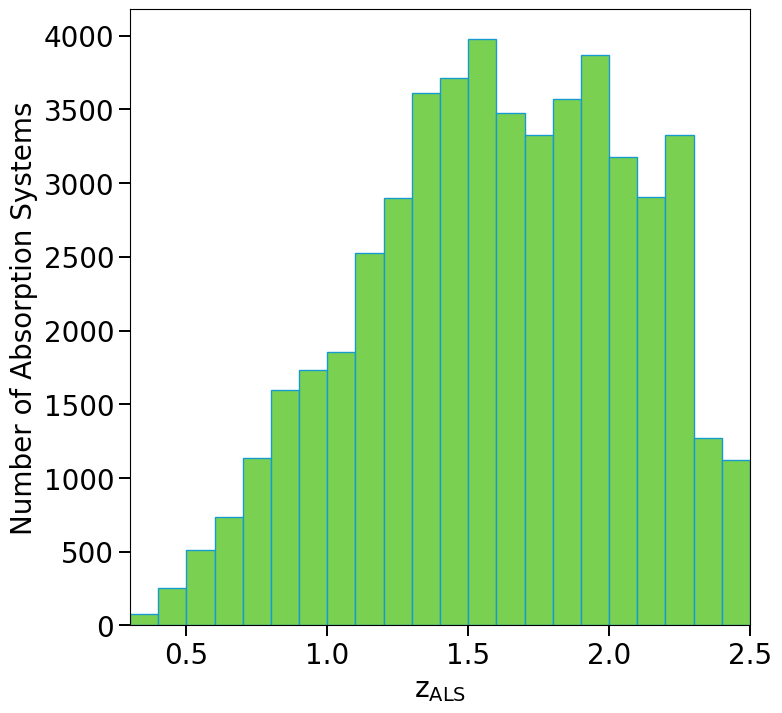}
\caption{Histogram of absorber redshift values for the 50{,}674 absorption line systems in our sample. Bins are 0.1 redshift wide.}
\label{ZABS}
\end{figure}

Following these cuts, our sample is reduced to 3{,}676 absorbers from the EDR and 47{,}720 from DR1. Of these 722 appear in both samples as the observations from these releases were not coadded. In all such cases, we choose to fit the EDR spectrum due to the higher data quality. This leaves our sample of QSO spectra with a single \Mgii\ absorber at \voff\ $ < 20{,}000$ \kms\ as having 50{,}674 entries. Figure \ref{ZABS} shows the distribution of absorber redshifts in this sample. We note that the number of absorbers rapidly increases from $z=0.3$ to $z=1.5$ which reflects the QSO redshift distribution in DESI \cite[e.g.][]{QSOTS}. At redshifts $1.5<z<2.3$ the distribution somewhat flattens, and beyond $z=2.3$ the number of detected absorbers significantly decreases. This is a result of the noise level of DESI spectra increasing at wavelengths greater than 8400\AA\ making the detection of \Mgii\ absorbers more challenging \citep{DESI_Instrument,MgIICat}.

In order to best understand the results of our extinction fitting process, we have additionally created a control sample of DESI QSOs with no detected \Mgii\ absorbers. In constructing this sample, we consider all $\sim1.266$ million QSOs in DESI DR1 that do not appear in the DR1 \Mgii\ catalog. We first trim this sample, removing those QSOs with g-band flux values less than the median value. We perform this cut as it is challenging to detect \Mgii\ absorbers in low flux QSO spectra \citep{MgIICat}, and we intend for our control sample to have similar properties to our absorbed QSO sample. We next match each absorbed QSO to a non-absorbed QSO by minimizing the difference in redshift, as well as signal-to-noise value. Once a non-absorbed QSO has been matched, we remove it from consideration for future matches. Following this procedure, we form our control sample 50{,}674 QSOs with no detected \Mgii\ absorbers. We will consider the results of our extinction fitting to this sample in \S3.

\subsection{Extinction Fitting}
In order to quantify the extinction resulting from our sample of absorption systems, we adopt methods similar to those used in \citet{Fawcett2022,Fawcett2023}. We first correct for Galactic extinction using the \citet{SFD} dust map and the \citet{MWCurve} Milky Way extinction law. We then resample each spectrum and its associated error spectrum into the rest frame of the absorption system using its redshift. For our control sample, we resample each spectrum according to the redshift of the absorption system detected in the QSO to which it was matched. To perform this shift we use the resample function as implemented in the {\tt Scipy.stats} package\footnote{https://github.com/scipy/scipy}, we note that this function uses Fourier methods, and so can be inaccurate at the edges of the resampled signal, as such we will mask these regions out of our fitting process \citep[][]{Scipy}. The wavelength grid we use in resampling has a spacing of 0.4\AA , roughly equal to the spacing of the DESI resolution element 0.8\AA\ shifted into the rest frame of an average absorber from our sample.

Fitting the amount of extinction present in each QSO requires we select both a blue QSO composite as well as an expected extinction curve. For our composite spectrum we employ the VLT/X-shooter control QSO composite from \citet{Fawcett2022}. This composite spectrum was constructed from 28 blue QSOs at redshifts $1.45<z<1.65$ which are unlikely to be significantly reddened based on their selection criteria, see \citet{Fawcett2022}. We choose to use this composite rather than construct our own from DESI spectra due to its wavelength coverage, in order to achieve similar coverage using DESI spectra we would need to use QSOs from a significantly wider redshift range, potentially resulting in a non-physical continuum shape.

\begin{table}[th!]
\centering
\caption{QSO Population Statistics}
\hspace{-45pt}
\begin{tabular}{c|c|c|c|c}
\hline
Sample & log(M$_\textrm{BH}$) & log($\lambda_\textrm{EDD}$) & log(L$_\textrm{BOL}$)  \\
    \hline
    \hline
    X-Shooter Blue\footnote{As described in \citet{Fawcett2022}}  & [8.4, 9.8] & [-1.5, 0.3] & [45.5, 47.0] \\
    DESI\footnote{Sample of 490,648 DESI DR1 QSOs from \citet{Pan2025}} & [7.9, 9.4] & [-2.0, 0.0] & [45.0, 46.0] \\
\label{tab:QSOChar}
\end{tabular}
\end{table}

\begin{figure*}[ht!]
\plotone{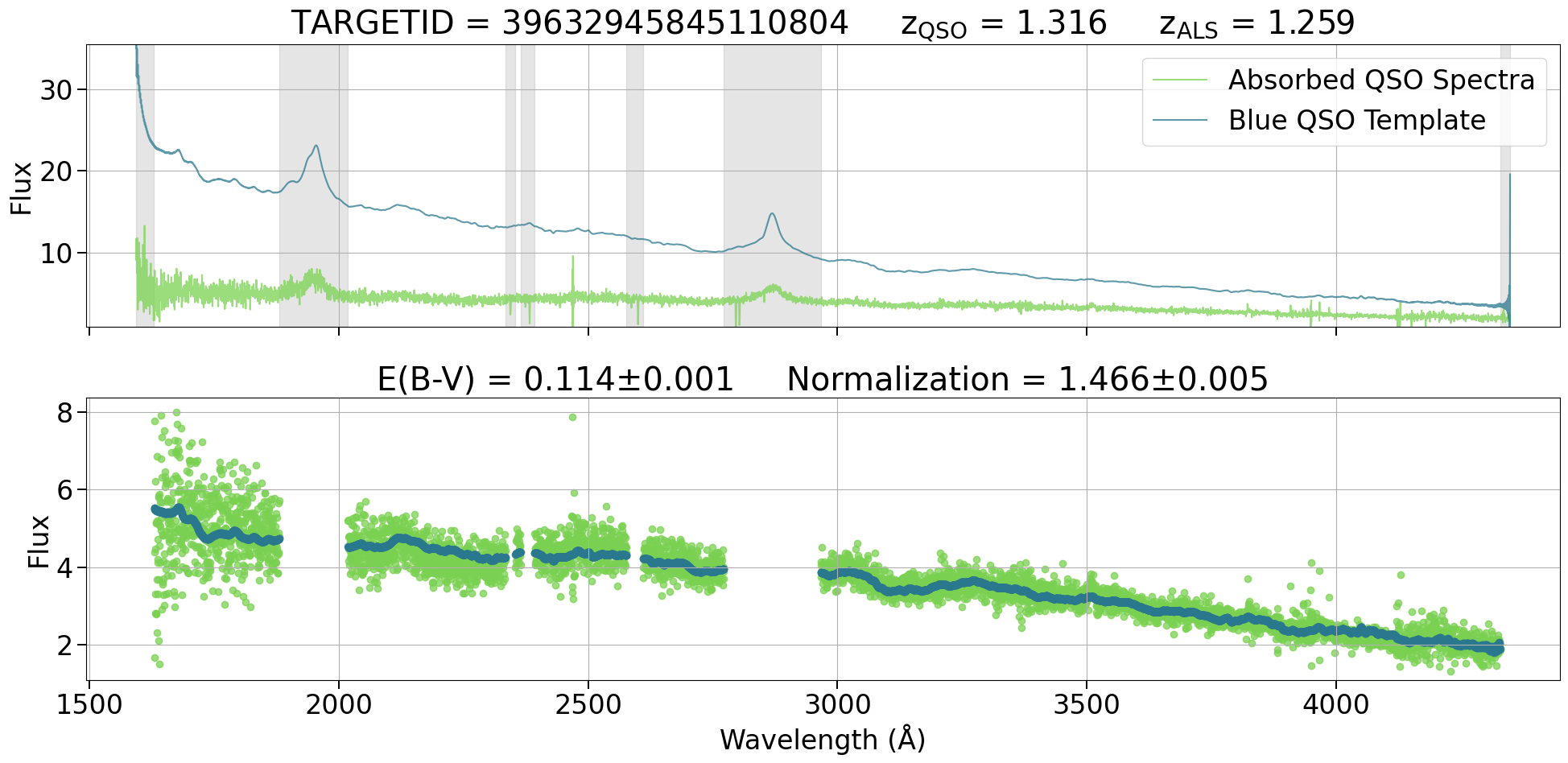}
\caption{Visualization of our extinction-fitting process. \textit{Top:} An example QSO spectrum with an absorber plotted alongside the blue QSO template, both shifted into the rest frame of the absorption system. Masked regions, including the edges of the spectrum as well as the emission and absorption lines, are shown in gray. Note that the template spectrum has been scaled arbitrarily. \textit{Bottom:} The result of the fitting process, showing that the reddened template with $\textrm{E(B-V)} = 0.114$ is a good fit to the unmasked data from the top panel.}
\label{fitprocess}
\end{figure*}

We can further consider the physical characteristics of the QSOs used in the construction of this composite, and compare these to our population of DESI QSOs. Table \ref{tab:QSOChar} presents the ranges of black hole masses, Eddington ratios, and bolometric luminosities for these QSOs as given in \citet{Fawcett2022}. We additionally provide the ranges of these same characteristics as determined in \citet{Pan2025} for a sample of DESI QSOs at redshifts $0.6 < z < 1.6$. Regrettably, these values have not yet been determined for the full sample of DESI QSOs, and so we are unable to give them for our full sample of absorbed QSOs. We note that the range of DESI QSO black hole masses given here are those calculated using the parameters from \citet{ShenBH}, as this same formulation of parameters was used in determining the X-shooter QSO black hole masses. From these data, we can note that the QSOs used in constructing the X-shooter composite tend to be brighter and associated with slightly more massive black holes when compared to the DESI population.

A further consideration is the overall shape and emission line widths of our blue composite spectrum. Considering the results of \citet{Fawcett2023}, particularly as shown in Figure 12, we can see that the X-shooter blue QSO composite spectrum is remarkable similar to the composite constructed using DESI QSOs with the lowest degree of reddening. Comparing these to the other composites plotted, we can see that the shape of the \Mgii\ emission line region is consistent.

As the X-shooter control composite is constructed in the rest frame before fitting, we first resample it such that it appears at the same wavelengths and with the same wavelength spacing as the absorbed QSO spectrum, as can be seen in Figure \ref{fitprocess}. Following the results of prior studies \citep[e.g.][]{York2008,ShenMenard2013,Khare2014,Chen2020} we adopt a Small Magellanic Cloud Bar extinction curve, with a ratio of selective-to-total extinction R(V) = 2.74. Specifically we adopt the results of \citet{SMCBar} as implemented in the {\tt Astropy} affiliated package {\tt dust\_extinction} \footnote{https://github.com/karllark/dust\_extinction/} \citep{Astropy,DustEXT}.

Before fitting we first mask a selection of pixels in each spectrum: the first and last 50 pixels at the edges of the wavelength space, which are often noisy due to resampling, pixels that fall within emission lines, as determined by eye using the blue-QSO composite, pixels that are within 10 \AA\ of the center of a significant absorption line\footnote{Specifically the Si IV, C IV, Fe II and Mg II absorption lines}, and pixels with errors values 5 standard deviations above the mean. Approximately 20\% of pixels in a given spectrum are affected by this masking. We then fit the blue QSO composite to the absorbed QSO spectra using a two parameter model, a normalization term and the color excess of the extinction curve. The color excess is given bounds $E(B-V) = [-1.0,2.5]$ and the normalization term is unbounded.

We note that the lower bound of $\textrm{E(B-V)} = -1.0$ would suggest a degree of reddening that is likely physically impossible for an absorption line system. Negative E(B-V) values require careful scrutiny and in general suggest that the DESI QSO being fit possesses a higher power-law slope than the utilized composite spectrum, which is possible given the natural spread in this value \citep[e.g.][]{Richards2003,Fawcett2023}, but surprising given the presence of an absorption system. By considering the \avgebv\ of a sample of absorbers, as we will in \S3, we should be able to effectively average over the range of possible power-law slopes and determine physically meaningful results. 

%As the parameters we will consider are those of the absorber and not the QSO, we would anticipate that they are independent of the power-law slope, and we find no evidence in the literature to the contrary.

Our fitting procedure is visualized in Figure \ref{fitprocess}. We perform this fitting for all 50{,}674 spectra in both our sample of absorbed QSOs, and our sample of control QSOs, and record the fit parameters and uncertainties. We will now consider the results of this process.

\section{Results}
\begin{figure}[ht!]
\plotone{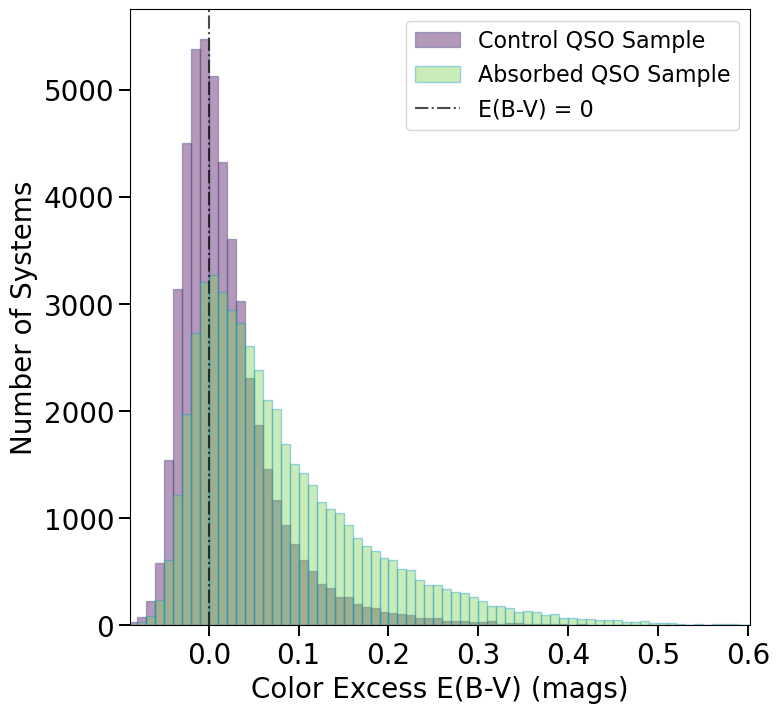}
\caption{Histogram of fit color excess values for all 50{,}674 spectra in both our sample of absorbed QSOs and our control sample of QSOs with no detected \Mgii\ absorbers. Bins are 0.01 magnitudes of extinction wide.}
\label{EBVHist}
\end{figure}

The distribution of $\textrm{E(B-V)}$ values we find for both our sample of absorbed and control QSOs is shown in Figure \ref{EBVHist}. We observe that for both samples, the most populated bins are those adjacent to $\textrm{E(B-V)} = 0$, and that the number of systems decreases in a consistent fashion towards more positive or negative values. The distribution of $\textrm{E(B-V)}$ values for our control QSOs is highly concentrated around $\textrm{E(B-V) = 0}$, whereas the distribution of $\textrm{E(B-V)}$ values for absorbed QSOs has a long tail of positive values. Quantifying these observations, we have calculated that the median extinction value for the absorbed QSO sample is: \avgebv\ $=0.052^{+0.120}_{-0.058}$, whereas the median value for the control sample is \avgebv\ $=0.007^{+0.055}_{-0.032}$, where the error bars given are the differences to the 16th and 84th percentile values in the respective distributions. From these results, we can see that the median absorbed QSO has a small reddening effect, whereas the median control QSO has essentially no reddening effect.

In order to gain a deeper understanding of the expected reddening effect of an individual absorber, we can consider its relationship with the other parameters of the absorber, as we will do in the following subsection. We will then consider the evolution of \voff\ values for associated absorbers with redshift and investigate their implications for QSO redshift values. Finally, we will outline a procedure for recalculating QSO redshifts that utilizes information about the absorption systems in an effort to understand and possibly reduce their redshift confusing effects. 

\subsection{\avgebv\ Trends}

\begin{figure}[ht!]
\epsscale{1.2}
\plotone{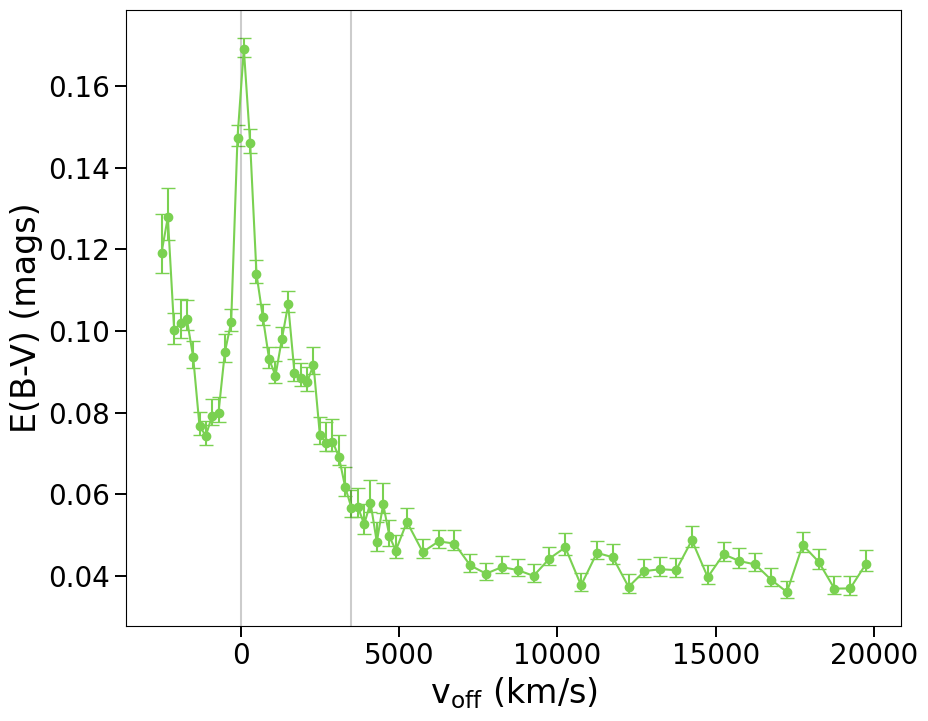}
\caption{\avgebv\ values in bins of \voff. Error bars are the standard error of the mean in each bin. Note that at \voff\ $<5000$ ${\rm km\,s}^{-1}$ we have drawn samples every 200 \kms, whereas at \voff\ $>5000$ ${\rm km\,s}^{-1}$ they are drawn every 500 \kms. Black lines are overlaid at \voff\ = 0 \kms\ and \voff\ = 3500 \kms.}
\label{EBV-VOFF}
\end{figure}

In order to investigate the delineation between associated and intervening absorbers we have plotted \avgebv\ as a function of \voff. This is shown in Figure \ref{EBV-VOFF}. We observe that \avgebv\ is relatively constant at \voff\ $>$ 5500 \kms. This behavior is relatively unsurprising as absorbers at these velocity offset values have large physical separation from the background QSO host galaxy. Considering absorbers between 3500 $<$ \voff\ $<$ 5500 \kms, we find that \avgebv\ values are slightly increased compared to those at greater \voff. However, this increasing trend is irregular and when considered alongside the uncertainties on these data make it challenging to cleanly separate these as a separate class of absorber. We make this observation because absorbers at these \voff\ values would have been considered associated systems by some older studies.

Below \voff\ $ = 3500$ \kms\ \avgebv\ increases rapidly and peaks at \voff\ $ = 0$\,\kms. This behavior is in sharp contrast to the behavior at \voff\ $ > 3500$ \kms\ and based upon these trends we will adopt the now commonly used value of 3500 \kms\ for the separation of associated and intervening absorption systems for the remainder of our analysis \citep[][]{VandenBerk2008,ShenMenard2013,Khare2014,Chen2020,Abhijeet2021}. Regarding the rapid rise at \voff\ $ < 3500$ \kms\ we can consider that these regions may be host to absorbers of different natures, as outlined in \S 1, and as such have differing cloud sizes and amounts of absorbing material.

Using 3500 \kms\ as our delineating value, we can note that the associated absorber sample contains 28{,}178 systems with \avgebv\ = 0.1092 magnitudes and the intervening absorber sample contains 22{,}496 systems with \avgebv\ = 0.0441 magnitudes.

The behavior at negative values of \voff\ is intriguing; we observe that the \avgebv\ decreases initially, and then increases again for the most negative \voff\ values. To better characterize the $\textrm{E(B-V)}$ trends for AAS we will need to consider its variation with redshift, as we do in Figure \ref{EBV_voff_zabs} and its discussion in section \S 3.2.

\begin{figure}[ht!]
\plotone{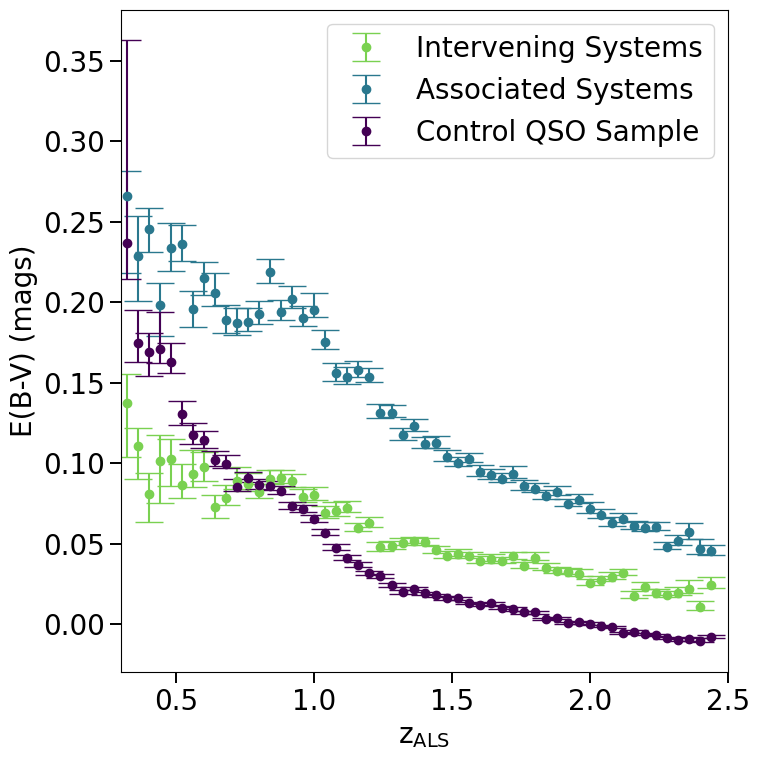}
\caption{\avgebv\ calculated for bins of absorber redshift. Bins are 0.04 units wide and the results are plotted for associated (\voff\ $<$ 3500 \kms) and intervening (\voff\ $>$ 3500 \kms) absorption systems, as well as for the sample of control QSOs.}
\label{ZABS-scale}
\end{figure}

Having now separated our samples of IAS and AAS, we can investigate the trends these samples, as well as our sample of control QSOs, show for \avgebv\, with respect to other parameters of the absorbers. Figure \ref{ZABS-scale} shows the relationship between absorber redshift and $\textrm{E(B-V)}$ values for these three samples. Note that the absorber redshift value assigned to our control QSOs is the same as the value of the absorbed QSO to which it was matched following the procedure described in \S2.1. Across all three considered samples, we observe that the color excess decreases for absorbers at higher redshifts. We find that this trend is more pronounced for associated systems when compared to intervening systems. This is in line with the understanding that galaxies become dustier over time as stellar populations evolve \citep[e.g.][]{DustyGal1,DustyGal3}. This trend being more pronounced in associated systems could also be reasonably predicted as QSOs are only found in relatively massive halos, which should primarily contain relatively massive galaxies \citep[e.g.][]{QSOHaloMass1,QSOHaloMass2,QSOHaloMass3}. 

Considering our sample of control QSOs, we observe that at redshifts z $ > 1 $ they consistently have E(B-V) values less than those of both associated and intervening systems. This is in line with our expectations for QSOs with no detected \Mgii\ absorbers. However, at redshifts z $ < 1$ we find that the E(B-V) of our control sample rises rapidly with a slope similar to the associated systems, and at z $<\ \sim 0.8$, possesses E(B-V) values greater than those of the intervening absorber sample. This result is unexpected, and we do not have a clear explanation for its occurrence. Potential explanations include the possibility that the low-z population of DESI QSOs without detected \Mgii\ absorbers are in fact quite reddened, or alternatively, that our control sample at these redshifts is poorly constructed, potentially due to the small number of DESI QSOs at these redshifts. It is also possible that this result could indicate issues with our fitting approach at these redshifts. This result demands further investigation, which we intend to pursue in future work.

%We suspect this may be related to a saturating of the reddening effect resulting from the increasingly dusty environments we would anticipate finding at low redshift.

\begin{figure}[ht!]
\plotone{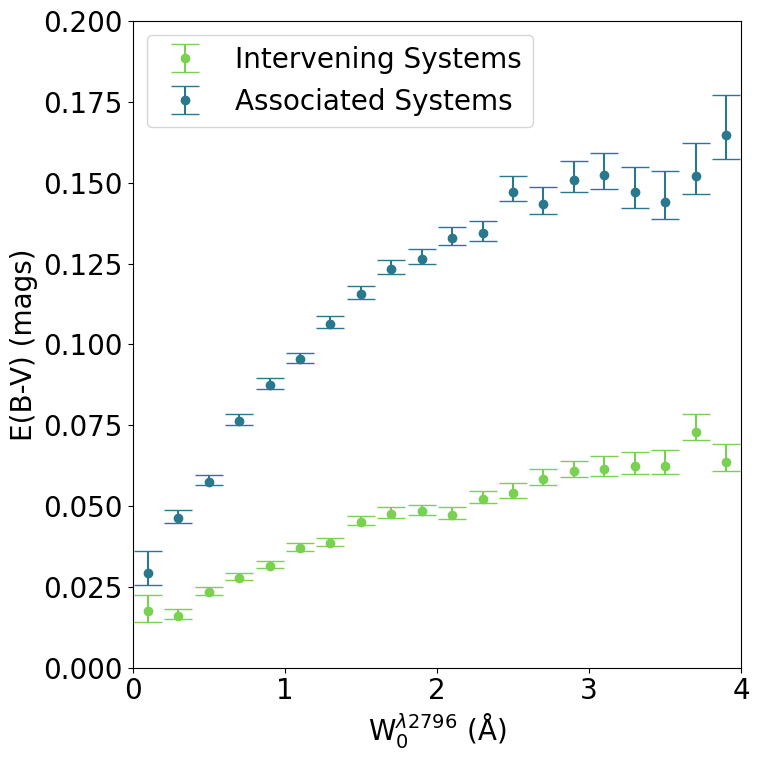}
\caption{\avgebv\ calculated for bins of \EWFL. Bins are 0.2 \AA\ wide and the results are plotted for both associated (\voff\ $<$ 3500 \kms) and intervening (\voff\ $>$ 3500 \kms) systems.}
\label{EW-scale}
\end{figure}

Figure \ref{EW-scale} shows the relationship between rest frame equivalent width \EWFL\ and $\textrm{E(B-V)}$ values. We do not plot our control QSO sample here, as the equivalent width of the absorption system to which our control entries were matched is not physically meaningful. We clearly observe that as \EWFL\ increases the \avgebv\ increases, at first rapidly, and then begins to level off at high values of \EWFL. \avgebv\ is greater at all \EWFL\ values for the associated systems compared to the intervening sample, and the rate at which \avgebv\ increases is greater as well. This clearly shows that \Mgii\ line strength and the degree of reddening an absorption system causes are connected. Relating the results of this plot to those seen in Figure \ref{ZABS-scale}, a self-consistent picture emerges wherein the absorbers with the greatest \EWFL\ values are those found at low redshift and both of these factors tend to result in a higher degree of reddening. 

\subsection{Redshift evolution of \voff\ distributions}

\begin{figure*}[ht!]
\plotone{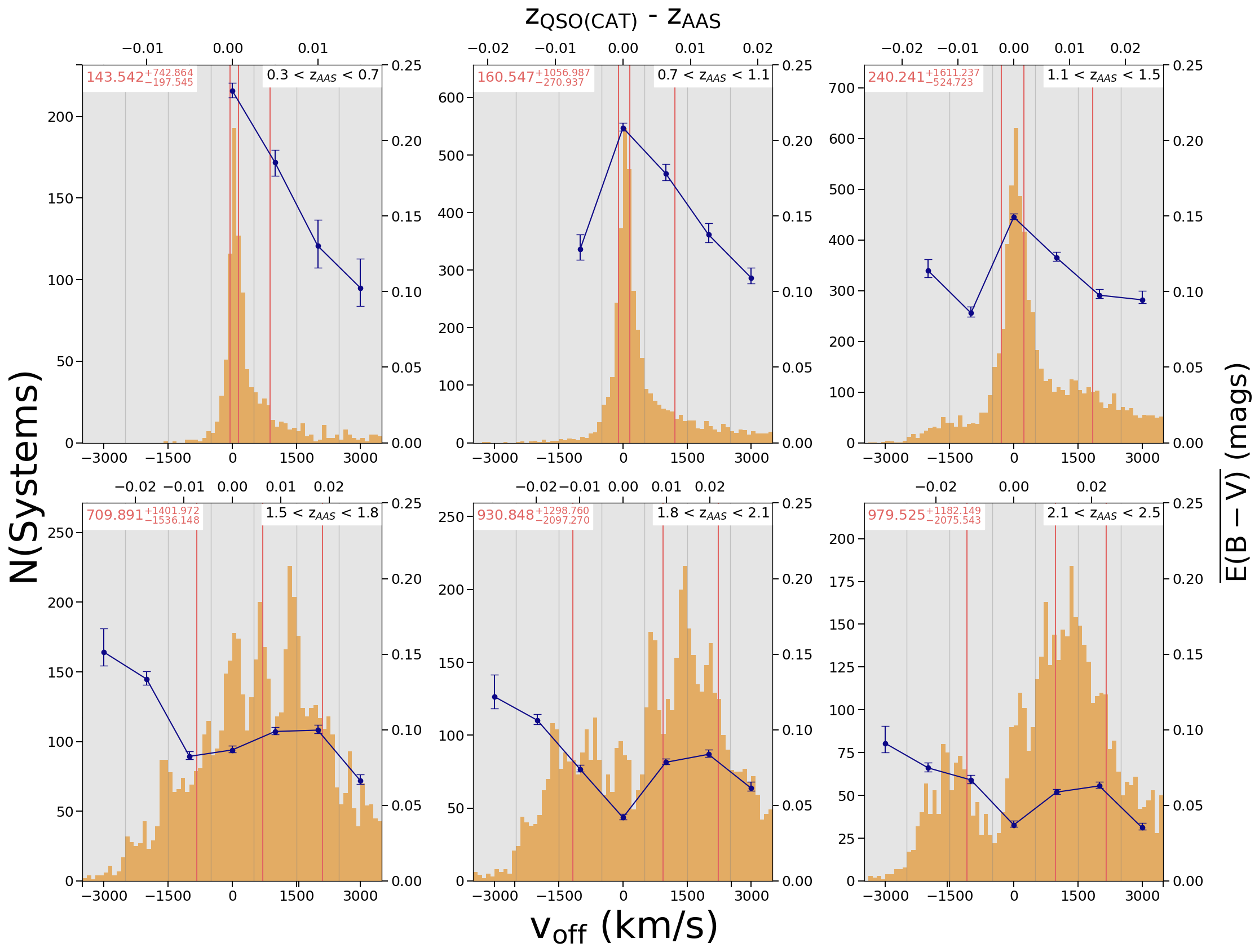}
\caption{Histogram distributions of \voff\ values, drawn every 100 \kms\ and shown in light orange, overlaid with \avgebv\ values, drawn every 1000 \kms\ and shown in dark purple, as calculated using redshift subsamples of associated absorbers. The redshift range of each panel is indicated in the upper right and the 50th percentile value of the \voff\ distribution, as well as the difference to the 16th and 84th percentile, is given in the upper left and these percentiles are displayed as light orange vertical lines. Note that the \avgebv\ values are calculated in seven 1000 \kms\ bins from -3500 \kms\ to +3500 \kms\ as indicated by the light gray boxes and are only shown if the corresponding \voff\ bin has at least 30 entries, error bars on these points are the standard error of the mean. The top axis of each panel has been converted from \voff\ to a difference in redshift using the mean redshift of each panel's redshift range.}
\label{EBV_voff_zabs}
\end{figure*}

We can now consider the evolution of \avgebv\ for associated systems as a function of redshift. As noted in \S 1, QSO redshifts can be imprecise due to their broad-line emission, and become more challenging to determine at high redshift \citep[e.g.][]{SDSS12}. This redshift imprecision translates into an uncertainty on the \voff\ of an associated absorber. In order to investigate how this may impact \avgebv\ we compare the distributions of \voff\ values and their \avgebv\ in bins of absorber redshift. We have chosen bins such that they span redshift regions of interest, however as a result they do not necessarily contain similar numbers of AAS. The result can be seen in Figure \ref{EBV_voff_zabs} which compares E(B-V) to voff.

We observe that at redshifts z $ < 1.5$ the majority of absorbers are found within \voff\ $ = \pm 500$ \kms\ and the \avgebv\ is highest in this bin. This is in line with the behavior we saw for the full sample in Figure \ref{EBV-VOFF} and our physical understanding that absorbers at the smallest \voff\ values are likely to be larger and dustier and therefore result in the highest degree of reddening. 

However, at redshifts z $ > 1.5$ we observe that the \voff\ distributions are significantly more broad and their peak has shifted away from \voff\ = 0 \kms. At redshifts z $ < 1.5$ the 50th percentile value of the \voff\ distribution is between 140-240 \kms\ whereas at z $ > 1.5$ this value is between 700-1000 \kms. Looking specifically at the two highest redshift bins, $1.8<$ z $ < 2.1$  and $2.1<$ z $ < 2.5$, we observe that the distribution has bifurcated and now has two distinct peaks located at $\sim \pm 1500$ \kms. 

This drastic evolution in \voff\ values with redshift lacks any clear physical motivation, and to further illustrate this we can consider the behavior of \avgebv\ in these redshift bins. At redshifts z $ < 1.5$ we observe that \avgebv\ has its highest value in the bin centered on \voff\ = 0 \kms\ and decreases at both lower and higher values. However, at z $ > 1.5$ the highest \avgebv\ is found at extreme negative values of \voff, and the bin centered on \voff\ = 0 \kms\ has among the lowest \avgebv\ in these samples.

\begin{figure*}[ht!]
\plotone{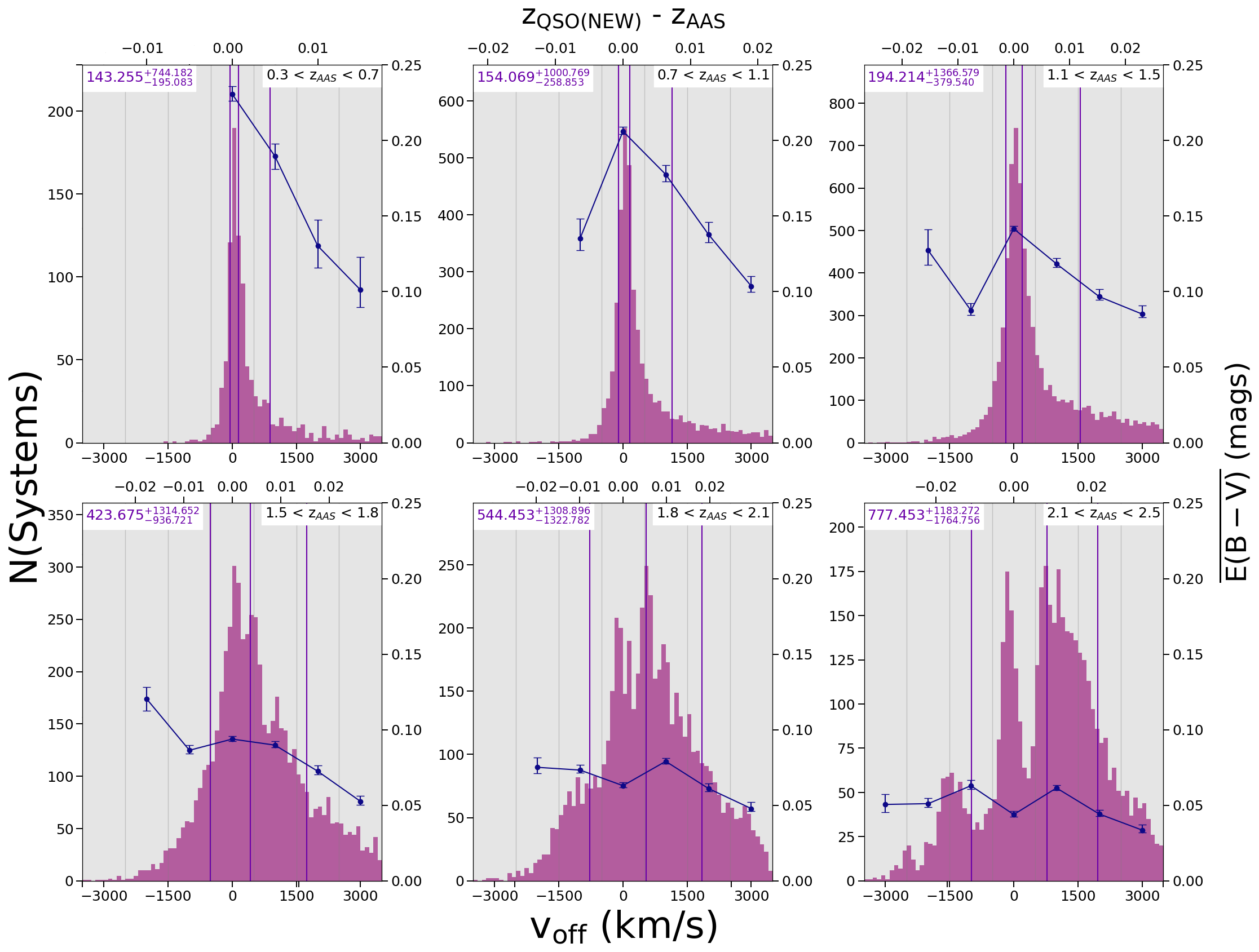}
\caption{A recreation of Figure \ref{EBV_voff_zabs} using \voff\ values as calculated from the z$_{\textrm{QSO}}$ values found when masking \Mgii\ absorption lines. The \voff\ histogram and vertical lines indicating the 16th, 50th and 84th percentile of each panel are now shown in light purple. Note that the \avgebv\ values, shown in dark purple, are calculated in seven 1000 \kms\ bins from -3500 \kms\ to +3500 \kms\ as indicated by the light gray boxes and are only shown if the corresponding \voff\ bin has at least 30 entries, error bars on these points are the standard error of the mean. The top axis of each panel has been converted from \voff\ to z$_{\textrm{QSO}}$ - z$_{\textrm{AAS}}$ using the mean redshift of each panel's redshift range. }
\label{ZDIFF-EVO}
\end{figure*}

Based upon these results, we strongly suspect that the presence of an associated \Mgii\ absorber is complicating the determination of QSO redshifts, particularly at z $ > 1.5$. As DESI QSO redshifts are determined by fitting templates to their spectra, we can consider that associated absorbers with the strongest absorption lines, which as we have seen from Figure \ref{EW-scale} result in the highest degree of reddening, will also have the greatest effect on the shape of \Mgii\ emission lines. The strong absorption lines may result in the \Mgii\ emission line splitting into two distinct peaks, which is a potential explanation for the bimodal behavior of the \voff\ distributions in the highest redshift bins in Figure \ref{EBV_voff_zabs}.

As we believe these lines are confusing redshift estimation, a possible remedy is to remove these lines from the QSO spectra and rerun {\tt Redrock} such that any possible bias from the presence of the \Mgii\ lines can be understood and potentially removed. In the next section we detail our methodology in attempting this procedure and describe the results we find.

\subsection{Recalculating z$_{QSO}$}

In order to mask out associated \Mgii\ absorption lines, we first collect the absorber redshift and line width information from our input \Mgii\ catalog for all 28{,}178 associated absorbers in our sample. Using these values we calculate which pixels are within a 5$\sigma$ Gaussian width of the line centers of either line of the \Mgii\ doublet. We then flag these pixels such that they will not be considered by {\tt Redrock} when performing redshift fitting.

A crucial consideration is the version of the {\tt Redrock} software used in calculating quasar redshifts. DESI's EDR was run with version 0.15.3, whereas DR1 used version 0.17.0. Since the publication of these releases, a new version of {\tt Redrock}, 0.20.0, has been made available that uses improved quasar templates derived from a sample $\sim400$ times larger than those used previously \citep{RRNewTemplates}. 

All future DESI releases are intended to use a version of {\tt Redrock} implementing  these new QSO templates. With these considerations in mind we have chosen to use version 0.20.0 of {\tt Redrock} with the updated QSO templates. In order to ensure we can separate the effect of our masking strategy from that of the new QSO templates, we have calculated redshifts for our sample of associated absorbers both with and without the masking of \Mgii\ absorption line pixels. 

In interpreting the results of these two sets of new redshift values we follow the decision matrix outlined in \citet{QSOTS}. We find that in the results for both the masked and non-masked re-runs, a small fraction of our sample returns a redshift value such that it would no longer be considered associated with the background QSO it was previously.

Intriguingly, we find that this is true for 784 absorbers when considering the re-run with \Mgii\ lines masked and true for 1052 absorbers when the lines were not masked. In some cases then, the masking of absorption lines has resulted in a QSO redshift value more in agreement with the original catalog value, despite the use of the updated QSO templates. As we are concerned with the effect of line masking, and not that of the new {\tt Redrock} templates, in this paper we choose to exclude such systems from our sample at present, although we may revisit them in future work.

We then adopt the redshift found from the masked re-run for all absorbers and using these new redshifts we determine new \voff\ values and plot the results in Figure \ref{ZDIFF-EVO}. Comparing to Figure \ref{EBV_voff_zabs}, we can see that the results below redshift z $ < 1.5$ have not been significantly changed. The distributions of \voff\ still have the same shape as when using the catalog redshifts and the \avgebv\ values follow the behavior as before. We do note however, that in the $1.1 < $ z $ < 1.5$ bin the number of systems with \voff\ $>\pm 750$ \kms\ has noticeably shrunk.

We do however see a drastic change in results at redshifts z $ > 1.5$ as the \voff\ distributions have narrowed and become more concentrated around \voff\ = 0 \kms, particularly in the $1.5 < \textrm{z}  < 1.8$ and $1.8 < \textrm{z}  < 2.1$ bins. The change in the width of the distributions can be clearly seen by the change in the 50th percentile value, which has decreased for all bins z $ > 1.5$ in comparison to the results of Figure \ref{EBV_voff_zabs}. We can also note that the number of systems at \voff\ $<$ 0 \kms\ has clearly shrunk as the 16th percentile values of these distributions have become less negative. However, these distributions are still clearly more broad than than those found at lower redshifts, and contain a significantly higher fraction of negative \voff\ absorbers.

Considering the \avgebv\ values, we observe a more subtle change than is apparent in the \voff\ values. The values are in better agreement with our physical expectations as well as the behavior at lower redshifts, i.e. the \avgebv\ in the bin containing \voff\ = 0 \kms\ has increased and the \avgebv\ values do not rise as steeply when considering systems with negative \voff. That the \avgebv\ values are not more in line with the expected behavior suggests that the QSO redshifts of these systems may be inaccurate and require further refinement. \citet{Fawcett2023} found that dusty QSOs observed by DESI were more likely to have incorrect redshifts. 

Notably, this approach seems to have been least effective for those systems in the highest redshift bin, $2.1 < $z $ < 2.5$, although we do note that the bins around \voff\ = 0 \kms\ are more populated than they were when considering the catalog redshifts, the distribution still has clear additional peaks at $\sim \pm 1500$ \kms. We attempted an additional {\tt Redrock} re-run where we additionally masked the approximate locations of the C IV and Si IV absorption lines and saw no significant improvement over the results of Figure \ref{ZDIFF-EVO} in this or any other redshift bin.

\begin{figure}[ht!]
\epsscale{1.2}
\plotone{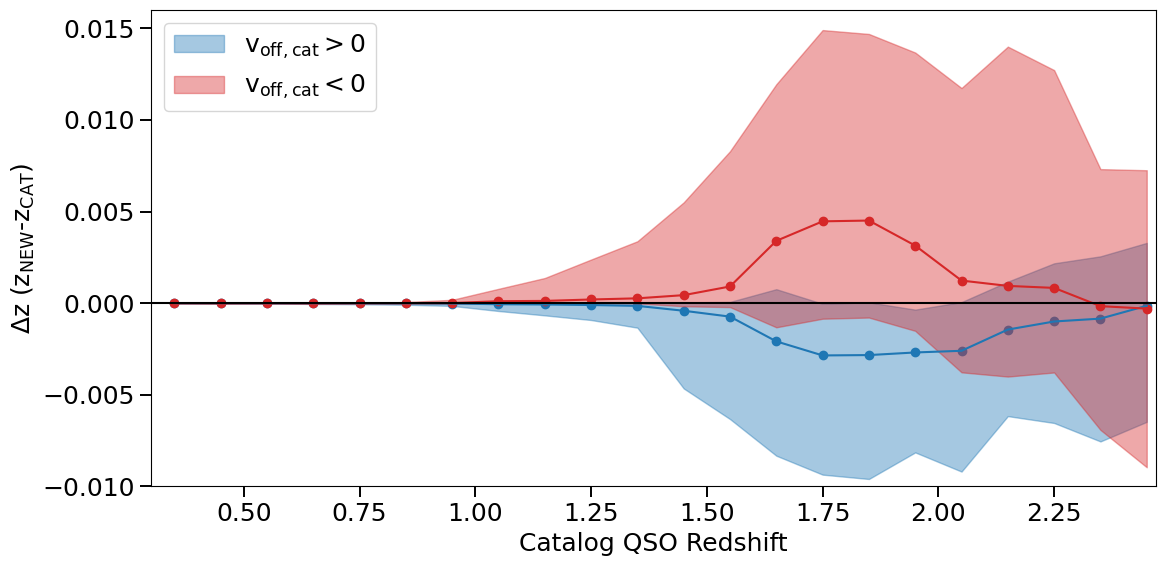}
\caption{The difference in redshift between the catalog values and the values found when the \Mgii\ lines are masked. $\Delta$z is calculated in redshift bins 0.1 units wide and the 50th percentile value is plotted with the surrounding colored envelope showing the 16th and 84th percentiles of the distribution. Results for systems initially at \voff\ $>0$ \kms\ are shown in blue and systems initially at \voff\ $<0$ \kms\ in red.}
\label{voff-diff}
\end{figure}

We have additionally plotted the average change in redshifts between the results shown in Figures \ref{EBV_voff_zabs} and \ref{ZDIFF-EVO} as a function of the initial catalog redshift, as seen in Figure \ref{voff-diff}. We have divided the sample into those systems initially redshifted or blueshifted relative to the background QSO, and in doing so it becomes clear that when the masking resulted in a change in the QSO redshift, the typical effect was to reduce the difference between the quasar and absorber redshift. This is in line with our observations that the distributions of \voff\ values narrowed when using the post-masking redshift values.

We note that there was essentially no change in redshift for systems at redshifts z $ < 1.0$. This could potentially be explained by the narrow QSO emission line O [III], which is visible in DESI spectra for redshifts z $< 0.85$, and which provides a particularly precise redshift estimate \citep[e.g.][]{BroadLineWidth}. At redshifts z $ > 1.0$ we find that the average change in redshift begins to increase and reaches its peak around redshift z = 1.8. At this peak the average $\Delta$z for QSOs with associated absorbers initially at negative \voff\ values is $\sim$0.005 and the top 16\% of $\Delta$z values are $\gtrsim$ 0.015. At redshifts z $ > 2.0$, the average $\Delta$z begins to decrease and at z = 2.5 is approaching 0. This finding is surprising, but is in line with the results of Figure \ref{ZDIFF-EVO} where our masking strategy seemed to be least effective in the highest redshift bin. We will discuss possible explanations for our masking strategy being less effective at these redshifts in \S4.

These results show clearly that the presence of an associated absorption system can significantly affect the redshift of a DESI QSO. This effect is most pronounced at redshifts 1.5 $<$ z $<$ 2.1 where the average $\Delta$z $\approx$ 0.005. Our results further suggest that these redshifts may require further refinement as even post-masking, the distributions of \voff\ values at z $>$ 1.5 are clearly distinct from those at lower redshifts, being more broad and having a significantly higher fraction of negative \voff\ values.

\section{Discussion}
The number of absorption systems found at \voff\ $< -1500$ \kms\ in our analysis is intriguing in regards to the physical conditions necessary to observe these systems. Such systems have been catalogued in small number in previous works and so are not novel \citep[e.g.][]{ShenMenard2013,Chen2020}. However, the results of Figures \ref{EBV_voff_zabs} and \ref{ZDIFF-EVO} show that these systems are exceedingly rare at redshifts z $ < 1.5$, and a number of those systems that were initially at \voff\ $< -1500$ \kms\ had less negative values after their redshift was re-calculated with the \Mgii\ absorption lines masked.

One consideration regarding these very negative \voff\ values, as discussed throughout this paper, is the uncertainty in redshifts determined using QSO broad lines. By studying the velocity shifts of various narrow and broad QSO emission lines from the SDSS \citet{BroadLineWidth} found that it is not feasible to measure QSO redshifts to better than an intrinsic uncertainty of $\sim 200$ \kms\ using only broad lines. This uncertainty would naturally broaden the expected distribution of \voff\ values, particularly at  redshifts z $\gtrapprox 1.6$ where narrow lines such as [O II] $\lambda$ 3727 $\textrm{\AA}$ and [Ne v] $\lambda$ 3426 $\textrm{\AA}$ become inaccessible in DESI spectra, and the broad lines with the greatest intrinsic uncertainties, C IV and Si IV, become accessible.

Another possible factor is the observed broadening of \voff\ distributions even at low redshift, where we believe the redshifts to be more accurately determined. This can be seen clearly in the first three plots of Figure \ref{ZDIFF-EVO}. Even when combining these factors, we struggle to propose a formation mechanism for these very negative \voff\ absorbers within their own galaxy environments. However, at redshifts z $> 1.5$ we would anticipate galaxies to be more actively accreting material from their surroundings. Perhaps then, these absorbers are associated with cold gas flows into these galaxy environments, like those expected to cease QSO activity \citep[e.g.][]{QSOQuench1,QSOQuench2,QSOQuench3,QSOQuench4}. Although in such a scenario it is unclear how the gas could have undergone sufficient enrichment to produce metal line absorption.

%We believe that this is unlikely to be explained by galactic dynamics as the typical mass of QSO hosting halos does not significantly grow with redshift \citep[e.g.][]{QSOHaloMass1,QSOHaloMass2,QSOHaloMass3}. Even combining these factors 

Relatedly, we can consider the effectiveness of our masking strategy at high redshift, as better performance in this regime could aid in determining if absorbers with very negative \voff\ values are real or a consequence of poorly determined redshifts. Recalling Figure \ref{voff-diff} where we see the average change in redshift decrease for absorbers at redshifts z $ > 2.0$, we can acknowledge the nature of DESI QSO spectra in this redshift regime. As noted above, a number of additional QSO broad emission lines, such as C IV, C III, and \LyA\ have become observable and are typically brighter than \Mgii. Additionally, at these redshifts \Mgii\ would be observed at $\lambda > 8000$\AA\ where the DESI spectrum tends to become noisier, and potentially obscured by skylines. Both of these factors would result in these other lines having greater signal-to-noise values and potentially being more influential in {\tt Redrock's} template fitting. In future work, we intend to explore possibilities to improve this performance by more careful masking of additional lines, as well as the possible use of other redshift fitting tools such as {\tt QuasarNet} \citep{QN}.

%Why not just use the absorber redshift as the QSO redshift? Spread in line/desi rr schema/etc

\section{Conclusion}
In this paper we have detailed the results of a study on the nature and effect of \Mgii\ absorption line systems on the spectra and redshifts of quasars observed by DESI. Using a sample of 50{,}674 Quasar spectra with a single detected absorption system at a velocity offset \voff\ $<$ 20000 \kms, as well as a control sample of 50{,}674 Quasars with no detected \Mgii\ absorbers, we have found that the median extinction value for the absorbed QSO sample is: \avgebv\ $=0.052^{+0.120}_{-0.058}$ and the median value for the control sample is \avgebv\ $=0.007^{+0.055}_{-0.032}$.

Considering the sample of absorbed quasars we have further found that the average color excess \avgebv\: increases rapidly at \voff\ $<$ 3500 \kms\ and tends to be greater for absorption systems at low redshift, as well as those with stronger \Mgii\ absorption lines, as parameterized by their equivalent width. These findings are largely in agreement with the idea that galaxies become dustier and more enriched with metals over time. Our finding that \avgebv\ tends to be greater for galaxies hosting associated absorbers at all redshifts would further suggest that these quasar hosting galaxies tend to be dustier when compared to galaxies hosting intervening absorbers.

Considering intervening systems, we find \avgebv\ to be consistently around 0.04 magnitudes at velocity offset values \voff\ $>$ 3500 \kms, affirming that these systems are independent of their background quasar. Considering Figures \ref{ZABS-scale} and \ref{EW-scale}, we note that while these systems share the behavior of associated absorbers, increasing with decreasing redshift and/or increasing \EWFL, they do so at a decreased rate, suggesting that the galaxies hosting these absorbers are less able to produce and/or retain dust and enriched elements.

In an effort to better understand the effect of associated \Mgii\ absorbers on the determined redshift of their QSO, we have investigated the evolution with redshift of \voff\ values found for these systems. Our results show that above redshift z $>$ 1.5, \voff\ distributions evolve in such a way that strongly suggests \Mgii\ absorbers are affecting redshift determination. Specifically, we find that the distribution drastically broadens and bifurcates, with absorbers becoming less common around \voff\ = 0 \kms\ and the absorbers with the highest \avgebv\ values being found at extremely negative \voff\ values. This behavior is in sharp contrast with that of absorbers at redshifts z $<$ 1.5 where the majority of absorbers, and particularly the absorbers with the greatest reddening effect, are found clustered around \voff\ = 0 \kms.

In order to demonstrate that this effect is caused by the presence of the \Mgii\ absorption lines we have re-run DESI's primary redshift determining tool, {\tt Redrock}, and masked those pixels that fall within the \Mgii\ lines as determined, using the line fitting information from our input catalog of absorbers. Our results show that post-masking the \voff\ distributions for absorbers at redshift z $>$ 1.5 are noticeably less bifurcated and absorbers have typically shifted back towards \voff\ = 0 \kms. This procedure is most effective for absorbers at redshifts between 1.6 $<$ z $<$ 2.0 where the typical change in QSO redshift is $\Delta z = 0.004$ with some redshifts being shifted by $\Delta z > 0.015$.

We find that the resulting \voff\ distributions for absorbers at redshifts z $>$ 1.5 are still not in full agreement with those at lower redshifts. This is most pronounced for absorbers at z $>$ 2.0, where this procedure seems to have been less effective. This is likely due to the general difficulty in estimating redshifts for quasars at z $ > 2$ where fits are dominated by broad lines.

Our results as presented here give additional insight into the host galaxy environments of both associated and intervening metal line absorbers and their evolution with redshift. Further we find that DESI QSO redshifts are significantly affected by metal line absorbers at z $>$ 1.5 and can be potentially improved by masking absorption lines in redshift fitting. We hope to improve upon the results of this paper in the future by employing a more sophisticated masking/redshift correcting strategy, utilizing more information regarding the absorption systems, incorporating additional redshift calculating tools, and by using a larger sample of absorbers. This will be possible using the upcoming DESI Year 3 sample or potentially the full five-year DESI sample of QSOs, which we expect to contain more than twice as many absorption line systems as the currently available DR1 sample \citep[e.g.][]{MgIICat}.

\begin{acknowledgments}
We would like to thank the anonymous referee for their constructive
comments that greatly helped improved the content and usefulness of
the paper.

LN and ADM were supported by the U.S.\ Department of Energy, Office of Science, Office of High Energy Physics, under Award Number DE-SC0019022.

This material is based upon work supported by the U.S. Department of Energy (DOE), Office of Science, Office of High-Energy Physics, under Contract No. DE–AC02–05CH11231, and by the National Energy Research Scientific Computing Center, a DOE Office of Science User Facility under the same contract. Additional support for DESI was provided by the U.S. National Science Foundation (NSF), Division of Astronomical Sciences under Contract No. AST-0950945 to the NSF’s National Optical-Infrared Astronomy Research Laboratory; the Science and Technology Facilities Council of the United Kingdom; the Gordon and Betty Moore Foundation; the Heising-Simons Foundation; the French Alternative Energies and Atomic Energy Commission (CEA); the National Council of Humanities, Science and Technology of Mexico (CONAHCYT); the Ministry of Science, Innovation and Universities of Spain (MICIU/AEI/10.13039/501100011033), and by the DESI Member Institutions: \url{https://www.desi.lbl.gov/collaborating-institutions}. Any opinions, findings, and conclusions or recommendations expressed in this material are those of the author(s) and do not necessarily reflect the views of the U. S. National Science Foundation, the U. S. Department of Energy, or any of the listed funding agencies.

The authors are honored to be permitted to conduct scientific research on Iolkam Du’ag (Kitt Peak), a mountain with particular significance to the Tohono O’odham Nation.
\end{acknowledgments}

The data used to construct the Figures in this paper are available on Zenodo at \dataset[DOI: 10.5281/zenodo.14166427]{https://doi.org/10.5281/zenodo.14166427}

\bibliographystyle{yahapj}
\bibliography{AbsEXT.bib}

\end{document}